\documentclass[aps,prl,twocolumn,longbibliography,10pt,superscriptaddress]{revtex4-2}
\usepackage{graphicx}
\usepackage{dcolumn} 
\usepackage{bm} 
\usepackage{braket}
\usepackage{color}
\usepackage{amsthm}
\usepackage{amssymb}
\usepackage{float}

\usepackage{comment}

\usepackage{tikz}

\usepackage{hyperref} 
\usepackage{qcircuit}       

\definecolor{lightblue}{RGB}{73,151,208}
\definecolor{crimson}{RGB}{140,41,53}

\hypersetup{
    colorlinks,
    linkcolor={crimson},
    citecolor={lightblue},
    urlcolor={lightblue}
}

\def\>{\rangle}
\def\<{\langle}

\usepackage{mathtools}

\begin{document}

\preprint{APS/123-QED}

\title{Fast Quantum Interconnects via Neutral Atom Ensembles} 

\author{Sina Zeytino\u glu }

\affiliation{
Institute of Theoretical Physics, TU Wien, Wiedner Hauptstraße 8-10/136, A-1040 Vienna, Austria }

\author{Wenchao Xu}

\affiliation{
Institute of Quantum Electronics, Department of Physics, ETH Z\"urich, Z\"urich 8093, Switzerland}
\affiliation{
Quantum Center, ETH Z\"urich, Switzerland}
\affiliation{
Laboratory for Nano and Quantum Technologies,
Paul Scherrer Institut, CH-5232 Villigen PSI, Switzerland}

\author{Thomas Pohl }

\affiliation{
Institute of Theoretical Physics, TU Wien, Wiedner Hauptstraße 8-10/136, A-1040 Vienna, Austria }

\begin{abstract}
The distribution of entanglement between distant qubits is a crucial element of scalable quantum computing. Here, we describe a scalable quantum interconnect that generates remote entanglement at rates approaching those compatible with two-qubit gates of current neutral-atom quantum processors. The proposed approach exploits the  strong dipole-dipole interactions between atomic Rydberg states to generate entanglement between stationary qubits and propagating photons, without the need for an optical cavity. We provide a thorough description of the optimal conditions for the developed entanglement-generation protocol for realistic experimental parameters  and demonstrate that entanglement-generation rates $\gtrsim 3\times 10^5$ s$^{-1}$ can be achieved using Rydberg-states of ytterbium atoms. Given the inherent scalability and design flexibility of the proposed interconnect, our results suggest a promising approach towards distributed networks based on neutral-atom quantum architectures.
\end{abstract}
\date{\today}%

\maketitle
The ability to share and transfer quantum information between spatially separate systems presents a necessary step towards scalable quantum technologies \cite{kimble2008quantum, wehner2018quantum,Awschalom2021}. Quantum interconnects that mediate interactions between distant nodes of a quantum network are indispensable for a wide range of applications from secure communication \cite{scarani2009security,renner2008security} and interferometric telescopes  \cite{gottesman2012longer} to networks of atomic clocks \cite{komar2014quantum,ludlow2015optical,nichol2022}, distributed quantum sensors \cite{ge2018distributed,zhuang2018distributed}, and modular quantum computation \cite{gottesman1999demonstrating,bravyi2022future,wu2024modular}. 
However, if the interconnects operate much slower than the typical operation times within the network nodes, they introduce a bottleneck that deteriorates the network performance \cite{wehner2018quantum,ang2024arquin,kozlowski2023rfc}.
Here, a key performance metric is the entanglement generation rate $\Gamma_{\rm e}=p/t_{\rm e}$, determined by the time $t_{\rm e}$ per entanglement-generation attempt and the probability $p$ to successfully generate the desired two-qubit state.
Fast interconnects, therefore, require strong and broadband coupling between photons and individual qubits.
\begin{figure}[t!]
    \centering
\includegraphics[width=0.96\columnwidth]{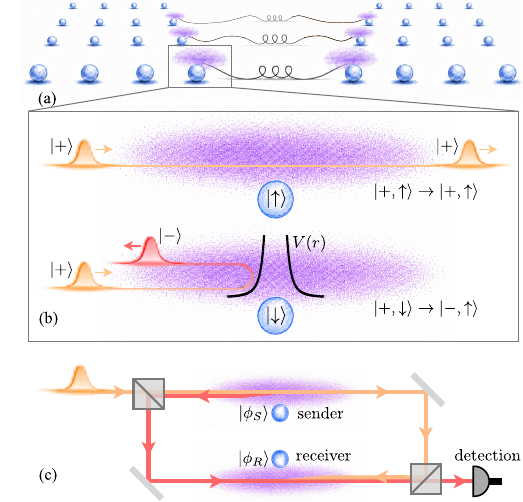}
    \caption{(a) Schematic illustration of a quantum interconnect for neutral-atom arrays that generates entanglement between atoms (blue spheres) in distant arrays by exploiting dipolar Rydberg-state interactions and collective photon coupling to mesoscopic atomic ensembles (purple clouds). (b) Dipolar exchange interactions, $V(r)$, between the qubit-atom and the propagating Rydberg-polaritons, that propagate through the ensemble, give rise to qubit-controlled photon transmission and reflection that generate entanglement between the qubit states, ($\lvert\uparrow\rangle$, $\lvert\downarrow\rangle$) and the propagation modes, ($\lvert\pm\rangle$), of the incident photon. (c) This conditional reflection and transmission can be used to deterministically entangle the states ($\lvert\phi\rangle_{S/R}$) of a sender and receiver qubit. The success of the entanglement generation protocol is heralded by photon-detection in a specific output mode. See text for details.
    }
    \label{fig:Fig1}
\end{figure}
Strong collective atom-light coupling can be achieved with dense atomic ensembles \cite{maxwell2013storage,heinze2013stopped,heshami2016quantum,busche2017contactless,wei2020broadband} or subwavelength atomic arrays \cite{adams2019rydberg,bettles2016enhanced,facchinetti2016storing, shahmoon2017cooperative,grankin2018free,rui2020subradiant,srakaew2023subwavelength}, while optical resonators are used to enhance the otherwise weak photon-coupling to individual atoms \cite{reiserer2015cavity,sipahigil2016integrated,humphreys2018deterministic,knaut2024entanglement}. 

Optical tweezer arrays of neutral atoms \cite{endres2016atom,barredo2016atom,cooper2018alkaline,norcia2018microscopic} have emerged among the leading platforms for quantum computing \cite{bernien2017probing,scholl2021quantum,bluvstein2021quantum,ebadi2022quantum,evered2023high,bluvstein2024logical}. Concurrently, the need for optical interconnects between neutral-atom processors has recently motivated substantial efforts \cite{huie2021multiplexed,young2022architecture,covey2023quantum,li2024high,li2025parallelized,sinclair2025fault,sunami2025scalable}. Cavity-based approaches are predicted \cite{li2024high} to generate entanglement at rates approaching $\Gamma_{\rm e}\sim10^{5} {\rm s}^{-1}$.
This would provide a much needed improvement of the current state-of-the-art \cite{li2025parallelized}, but remains below the achievable speed of rapid two-qubit gates that only require a few $100$ns \cite{evered2023high}.
Increasing the entanglement generation rate $\Gamma_{\rm e}$ is possible through parallel operation \cite{li2024high,sunami2025scalable} of cavity or nano-photonics setups but remains technologically challenging \cite{xu2025dynamics}.  

Here, we describe and analyze a different approach for fast and scalable quantum interconnects by utilizing the collective light-matter coupling in small atomic ensembles. The proposed method leverages the key resource of neutral-atom quantum processors, namely the strong dipole-dipole interaction between highly excited Rydberg states \cite{saffman2010quantum}. The resulting interaction between a single qubit atom and an adjacent ensemble is used to achieve strong coupling between the qubit and the mode-encoded quantum states of individual photons that propagate through the atomic ensemble [see Fig.\ref{fig:Fig1}(b)]. We outline a protocol that utilizes this mechanism for generating entanglement between two remote qubits. Numerical simulations of the qubit-photon dynamics are performed to optimize the proposed setup and reveal simple scaling laws of the entanglement-generation rate with key parameters. For the specific example of ytterbium atoms \cite{saskin2019narrow,jenkins2022ytterbium,ma2023high,nakamura2024hybrid,norcia2024iterative}, we show that entanglement rates in the range of $\Gamma_{\rm e}\sim 3\times 10^5$s$^{-1}$ are possible under achievable experimental conditions \cite{peyronel2012quantum, schlagmuller2016probing,das2025quantum}, with parallel operation of a handful of interconnects [Fig.~1(a)] extending this to aggregate rates approaching $\sim 10^6$ s$^{-1}$. Given the experimental feasibility of the key underlying mechanisms, such as dipole-exchange between Rydberg-polaritons \cite{thompson2017symmetry,busche2017contactless} and atomic qubit-ensemble coupling \cite{xu2021fast}, combined with the scalability of optically trapped ensembles, the proposed approach offers a viable and promising route to fast quantum interconnects for neutral-atom quantum processors.  

\begin{figure}
    \centering
    \includegraphics[width=1.\linewidth]{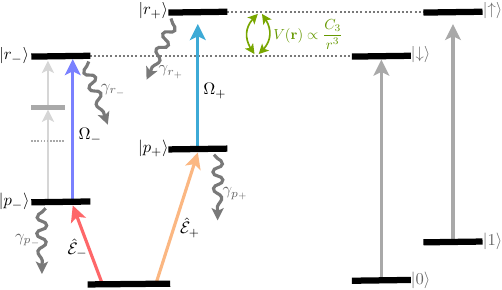 }
    \caption{Level diagrams for the ensemble atoms (left) and qubit atom (right). The coupling of the forward and backward propagating photon modes ($\hat{\mathcal{E}}_\pm$), the control-field Rabi frequencies ($\Omega_\pm$), and the decay rates $\gamma_{p_\pm}$ and $\gamma_{r_\pm}$ are indicated in the figure. The dipole-dipole interaction potential, $V({\bf r})$ between the Rydberg qubit and the ensemble atoms induces an excitation-exchange, $\lvert r_-,\uparrow\rangle\leftrightarrow \lvert r_+,\downarrow\rangle$, as described by Eq.(\ref{eq:Interactions}).}
    \label{fig:Fig2}
\end{figure}

Each node of our interconnect consists of an atomic qubit that is positioned close to an atomic ensemble \cite{vsumarac2026controlling}, as illustrated in Fig.\ref{fig:Fig1}(a) and (b). Incident photons in two counter-propagating modes, described by the field operators $\hat{\mathcal{E}}_+({\bf r})$ and $\hat{\mathcal{E}}_-({\bf r})$, can enter the ensemble from two opposite sides in respective propagation modes $\lvert+\rangle$ and $\lvert-\rangle$ and couple the atomic ground state to respective excited states $\lvert p_\pm\rangle$, as shown in the level diagram in Fig.\ref{fig:Fig2}. The state $\lvert p_+\rangle$ is coupled to a high lying Rydberg state $\lvert r_+\rangle$ via a classical control field with Rabi frequency $\Omega_+$, while $\lvert p_-\rangle$ is coupled to a Rydberg state $\lvert r_-\rangle$ with opposite parity via a two-photon transition with an effective Rabi frequency $\Omega_-$ \cite{supplementary}. Photon losses from spontaneous decay of the excited states with respective rates $\gamma_{p_{\pm}}$ and $\gamma_{r_\pm}$ are suppressed by operating at two- and three-photon resonance under conditions of electromagnetically induced transparency (EIT) \cite{fleischhauer2005electromagnetically}, giving rise to the slow-light dynamics of two counter-propagating Rydberg-dark-state polaritons \cite{fleischhauer2005electromagnetically,pritchard2010cooperative,gorshkov2011photon, 
petrosyan2011electromagnetically,dudin2012strongly,firstenberg2013attractive,murray2016quantum,firstenberg2016nonlinear,roy2017colloquium,thompson2017symmetry}.


The qubit atom, placed in close proximity to the ensemble, features two stable low-lying qubit states $\lvert0\rangle$ and $\lvert1\rangle$ that can be optically transferred to respective Rydberg states $\lvert\downarrow\rangle$ and $\lvert\uparrow\rangle$ (see Fig.\ref{fig:Fig2}), which are also of opposite parity. Choosing the $\lvert\downarrow\rangle\!\leftrightarrow\!\lvert\uparrow\rangle$ transition energy resonant with the $\lvert r_-\rangle\!\leftrightarrow\!\lvert r_+\rangle$ transition of the ensemble atoms, the Rydberg-qubit undergoes a resonant dipolar exchange interaction, described by 
\begin{align}
\hat{V} = \sum_j \,\, V({\bf R}-\mathbf{r}_j)\,\left(\ket{\downarrow}\bra{\uparrow} \otimes \ket{r_+}_j \bra{r_-}+ {\rm h.c.} \right),
\label{eq:Interactions}
\end{align} 
where $V(\mathbf{r})$ is the dipole-dipole interaction potential, whose strength is determined by the corresponding $C_3$ coefficient \cite{ravets2015measurement}. The position of the qubit atom is denoted by ${\bf R}$ and ${\bf r}_j$ is the position of the $j^{\rm th}$ atom in the ensemble. Upon matching the phase of the $\lvert r_\pm\rangle$-Rydberg spinwaves by suitably aligning the control-laser fields \cite{supplementary}, the dipole-dipole interaction can be used to mediate a transfer of the photon states between the $\lvert+\rangle$- and $\lvert-\rangle$-modes, conditioned on the state of the qubit. More concretely, we aim to implement the following operations 
\begin{align}\label{eq:mapping}
\lvert+\downarrow\rangle\rightarrow\lvert-\uparrow\rangle&\:,\quad\lvert+\uparrow\rangle\rightarrow\lvert+\uparrow\rangle,\nonumber\\
\lvert-\downarrow\rangle\rightarrow\lvert-\downarrow\rangle&\:,\quad\lvert-\uparrow\rangle\rightarrow\lvert+\downarrow\rangle.
\end{align}
This reflection or transition of the incident photon controlled by the qubit atom  (see Fig.$\ref{fig:Fig1}$) provides the basic mechanism for the two-qubit entanglement protocol. 

We analyze this process by numerically simulating the dynamics of the field amplitudes $E_{\pm,\sigma}({\bf r},t)$ for the photon propagation through the ensemble. Here, the slowly varying light-field amplitudes are normalized such that $|E_{\pm,\sigma}({\bf r},t)|^2$ yields the time-dependent probability density to find a forward ($\lvert+\rangle$) or backward ($\lvert-\rangle$) propagating photon at position ${\bf r}$ with the qubit in state $\lvert\sigma\rangle=\lvert\downarrow\rangle$ or $\lvert\sigma\rangle=\lvert\uparrow\rangle$. Starting from the full set of evolution equations for the photon modes and all atomic spinwave amplitudes \cite{supplementary}, one obtains a simple set of optical propagation equations in frequency space $\partial_z \tilde{\mathbf{E}}(\mathbf{r},\omega) = M(\mathbf{r},\omega)\tilde{\mathbf{E}}(\mathbf{r},\omega)$, where \cite{supplementary} 
\begin{align}
M &\!\equiv\! \left(\!\begin{array}{cccc} \chi_0(\mathbf{r},\omega) & 0 & 0 & 0 \\0 & \chi(\mathbf{r},\omega) & i\mathcal{V}(\mathbf{r},\omega)& 0 \\ 0 & -i\mathcal{V}(\mathbf{r},\omega) & -\chi(\mathbf{r},\omega)& 0\\
0&0 &0 & -\chi_0(\mathbf{r},\omega)
\end{array}\! \right),
\label{eq:MainResponse}
\end{align} 
and the vector $\tilde{\mathbf{E}}= \left(\tilde{E}_{+,\uparrow},\tilde{E}_{+,\downarrow},\tilde{E}_{-,\uparrow},\tilde{E}_{-,\downarrow}\right)^{\intercal}$ contains the relevant Fourier-transformed light-field amplitudes. Here, $\chi_0({\bf r},\omega)$ corresponds to the optical susceptibility of the medium in the absence of Rydberg-state interactions, i.e., for the input states $\lvert+\uparrow\rangle$ and $\lvert-\downarrow\rangle$ in Eq.(\ref{eq:mapping}). The corresponding amplitudes determine the transmission spectrum $T_0(\mathbf{r}_{\perp},\omega)\equiv \frac{\tilde{E}^{\rm (out)}_{+, \uparrow}(\mathbf{r}_{\perp},\omega)}{\tilde{E}^{\rm (in)}_{+,\uparrow}( \mathbf{r}_{\perp},\omega)}$ \cite{supplementary} of the ensemble at the transverse position $\mathbf{r}_{\perp}$, with $\tilde{E}^{\rm (in)}_{+,\uparrow}( \mathbf{r}_{\perp},\omega)$ being the incident field mode amplitude at some longitudinal position $z$ outside the ensemble while $\tilde{E}^{\rm (out)}_{+,\uparrow}( \mathbf{r}_{\perp},\omega)$ denoting the outgoing field following the propagation under Eq.(\ref{eq:MainResponse}). The coefficients $\chi$ and $\mathcal{V}$ determine the conditional reflection spectrum $R(\mathbf{r}_{\perp},\omega) \equiv \frac{\tilde{E}^{\rm (out)}_{-,\uparrow}( \mathbf{r}_{\perp},\omega)}{\tilde{E}^{\rm (in)}_{+,\downarrow}(\mathbf{r}_{\perp},\omega)}$ in the presence of exchange interactions, i.e. for the initial states $\lvert+\downarrow\rangle$ and $\lvert-\uparrow\rangle$ in Eq.(\ref{eq:mapping}). For simplicity, we consider here symmetric configurations with $\Omega_{\pm} = \Omega$, $\gamma_{p_{\pm}}=\gamma_p$ and $\gamma_{r_{\pm}}=\gamma_r$, but generalizing Eq.(\ref{eq:MainResponse}) is straightforward. Neglecting Rydberg-state decay ($\gamma_r=0$) and in the limit of long input pulses ($\omega\approx0$), the coefficients assume a particularly simple form
\begin{align}
    \nonumber \chi(\omega = 0, \mathbf{r})&\approx -\frac{\nu(\mathbf{r})}{l_{\rm abs}} \frac{U^2(\mathbf{r})}{1+U^2(\mathbf{r})},\\ 
    \mathcal{V}(\omega = 0, \mathbf{r})&\approx -\chi(\omega=0,\mathbf{r})/U(\mathbf{r}), 
\end{align}
while the full expressions are given in \cite{supplementary}. Here, $l_{\rm abs}\equiv \frac{g^2}{c\gamma_p}$ is the absorption length of the ensemble and $\nu(\mathbf{r})$ describes the dimensionless profile of the atomic density, with $\nu({\bf 0})=1$ at the center of the cloud. The strength of the dimensionless interaction potential $U(\mathbf{r})= \frac{V(\mathbf{r})}{\Omega^2/\gamma_p}\propto (r/r_b)^{-3}$ is determined by the Rydberg blockade radius $r_b=[C_3/(\Omega^2/\gamma_p)]^{1/3}$ that depends on the $C_3$-coefficient of the dipole-dipole interaction and the width, $\Omega^2/\gamma_p$, of the EIT window. 

The setup for the proposed entanglement generation protocol is illustrated in Fig.\ref{fig:Fig1}(c) and uses a sender-receiver topology \cite{jones2016design,beukers2023tutorial,supplementary} to generate the two-qubit Bell state
\begin{align}\label{eq:target_state}
\lvert\psi\rangle=\frac{\ket{0}_{S}\ket{1}_{R}+\ket{1}_{S}\ket{0}_{R}}{\sqrt{2}}
\end{align}
via cascaded interactions of the incident photon with the two qubit-coupled ensembles.
The sender qubit is initialized in $\ket{\phi}_{S} = \frac{1}{\sqrt{2}}(\ket{0}_{S}+\ket{\downarrow}_{S})$ and the receiver qubit is prepared in $\ket{\phi}_{R} = \ket{\downarrow}_{R}$. As the flying qubit enters the sender ensemble in the $\lvert+\rangle$-mode, it is either transmitted or reflected for the $\ket{0}_{S}$- or $\ket{\downarrow}_{S}$-component of $\ket{\phi}_{S}$, and, thereby, becomes entangled with the sender qubit. Afterwards, the photon enters the receiver ensemble from opposite sides, depending on the sender state. It induces a Rydberg-state transition upon reflection conditioned on the $\lvert0\rangle_S$ component of the sender atom. This second conditional reflection entangles the two distant qubits, whereby the success of Bell-state generation can be heralded via subsequent measurements. To this end, one first post-selects on the detection of the photon in the $\lvert-\rangle$-output port of the receiver ensemble [see Fig.\ref{fig:Fig1}(c)] at a time $t_d$ and position ${\bf r}_d$. Subsequently, both atoms are optically de-excited from the $\lvert\uparrow\rangle$-Rydberg state to their long-lived $\lvert1\rangle$ qubit state, while mapping the $\lvert\downarrow\rangle$-Rydberg state of the receiver atom to $\lvert0\rangle_R$. Upon probing the Rydberg-state population of the sender atom and discarding all events in which it is detected in a Rydberg state, the two-body state of the distant qubits is projected onto 
\begin{align}
    \ket{\psi}\! \propto\!  E_{T_0R}({\bf r}_d,t_d)\ket{0}_S\ket{1}_R + E_{RT_0}({\bf r}_d,t_d)\ket{1}_S\ket{0}_R.
    \label{eq:PostMeas}
\end{align}
Here, $E_{T_0R}({\bf r},t)$ denotes the output amplitude of the photon following transmission through the sender ensemble and reflection at the receiver node and vice versa for $E_{RT_0}({\bf r},t)$. For identical ensembles at the sender and receiver node and identical propagation phases acquired along the two paths, one has $E_{T_0R}=E_{RT_0}$ and, thus, generates the Bell state Eq.(\ref{eq:target_state}) with unit fidelity. 

Considering photodetection in any spatial mode, the corresponding success probability is obtained from \cite{supplementary}
\begin{align}\label{eq:ps}
p&=\!\int\!{\rm d}{\bf r}\: |E_{T_0R}({\bf r},t_d)|^2\nonumber\\
&=\!\int\!{\rm d}{\bf r}_\perp{\rm d}\omega\: |T_0(\mathbf{r}_{\perp},\omega)|^2|R(\mathbf{r}_{\perp},\omega)|^2|\tilde{E}^{\rm (in)}(\mathbf{r}_{\perp},\omega)|^2,
\end{align}
for a given spatio-temporal mode $\tilde{E}^{\rm (in)}(\mathbf{r}_{\perp},\omega)$ of the incident photon. From Eq.(\ref{eq:ps}) we obtain the entanglement-generation rate $\Gamma_{\rm e}=p/t_{\rm e}$, where the time $t_{\rm e}$ for each entanglement-generation attempt is composed of the duration of the output pulse and the total time delay from passing through the two ensembles \cite{supplementary}. Longer pulses typically yield a higher success probability $p$, while shorter pulses allow for shorter repetition times $t_{\rm e}$ at the cost of reducing $p$. This interplay yields an optimal pulse duration that maximizes $\Gamma_{\rm e}$ for a given geometry and set of parameters.

\begin{figure*}[t!]
    \centering    \includegraphics[width=\linewidth]{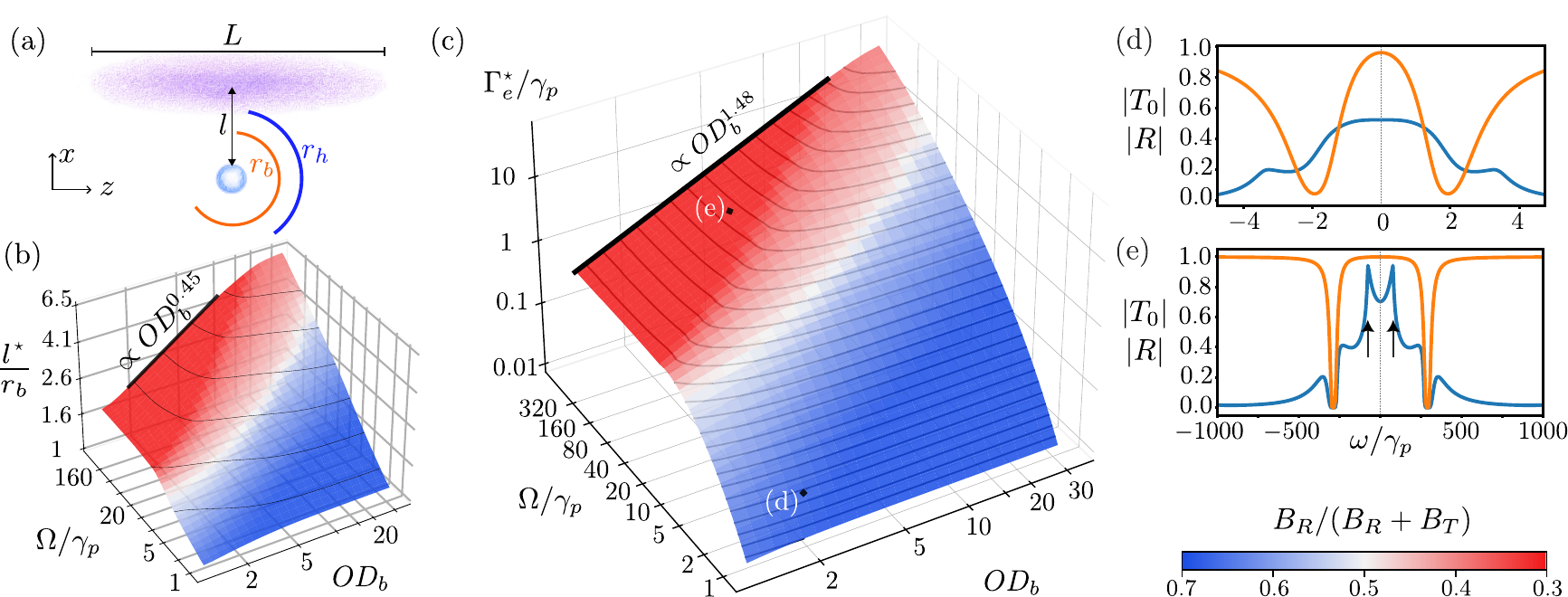}
    \caption{Optimizing the geometry of the setup (a) yields the optimal ensemble-qubit distance shown in panel (b), with the indicated scaling $l^\star\sim OD_b^{0.45}$ (black line) at large Rabi frequencies $\Omega$. Panel (c) shows the maximum entanglement generation rate for optimized geometry and pulse length as a function of $OD_b$ and $\Omega$. The equipotential lines correspond to constant values of $\Gamma_{\rm e}^\star$, while the color coding indicates the relative importance of the bandwidth for transmission ($B_T$) and reflection ($B_R$), according to $B_R/(B_R+B_T)$, with the colors saturating outside the range indicated in the legend. Blue shading marks indicate the transmission-limited regime, where the transmission-bandwidth in the absence of interactions [see the orange $|T_0|$ profile in panel (d)] is smaller than the bandwidth of reflection [see the blue $|R|$ profile in panel (d)], induced by the dipole-dipole interaction. Red shading in panel (c) corresponds to the opposite limit that is illustrated exemplarily in panel (e). The two vertical arrows mark the position of the reflection resonances that determine the reflection bandwidth and give rise to the high-intensity $\Gamma_{\rm e}^\star\sim OD_b^{3/2}$ scaling, indicated by the black line in panel (c).  }
    \label{fig:Fig3}
\end{figure*}

Taking into account Gaussian input pulses and Gaussian atomic density profiles for the two ensembles, the resulting entanglement-generation rate is determined by 5 parameters: the length $L$ of the ensemble, its lateral distance $l$ from the qubit [see Fig.$\ref{fig:Fig3}$ (a)], the scaled control field Rabi frequency $\Omega/\gamma_p$, the Rydberg-state decay rate $\gamma_r/\gamma_p$, and the optical depth $OD_b=r_b/l_{\rm abs}$ per blockade radius. For an efficient optimization of these parameters, we first discard transverse-mode effects and solve the effectively one-dimensional photon propagation along the $z$-axis for ${\bf r}_\perp={\bf 0}$ \footnote{Such an effective 1D treatment is often applied for simulations of Rydberg-polaritons \cite{fleischhauer2005electromagnetically,thompson2017symmetry}}. 

Figures \ref{fig:Fig3} (d-e) show representative transmission and reflection spectra from which different parameter regimes can be identified. For small values of $\Omega/\gamma_p$, $T_0(\omega)$ [Fig.\ref{fig:Fig3} (d)] displays the typical EIT features with near-perfect transmission on two-photon resonance ($\omega=0$) and a width $\Omega^2/\gamma_p$ of the transparency window, while the reflection spectrum shows a similarly simple structure with a broader frequency width but lower overall reflection amplitude. Hence, in this low-intensity regime, $t_{\rm e}$ is limited by the bandwidth, $B_T$, of photon transmission. 
Upon increasing the control-field Rabi frequency, the transmission spectra enter the Autler-Townes regime of slow-light propagation \cite{autler1955stark,holloway2014sub, saglamyurek2018coherent}. In this case, the transmission window flattens around $\omega=0$ while its width now scales linearly as $\Omega$. More importantly, the reflection coefficient develops a more complex spectral form with two pronounced maxima close to $\omega=0$, marked by the vertical arrows in Fig.\ref{fig:Fig3}(e). This pronounced double-peak structure arises from the strong dipole-dipole interaction and determines the bandwidth, $B_R$, of photon reflection. For sufficiently large values of $OD_b$, the near-resonant transmission and reflection amplitudes remain close to unity, such that their spectral bandwidths, $B_T$ and $B_R$, present the limiting factors of $\Gamma_{\rm e}$, as we describe below. 
 
The optimal qubit distance $l^\star$ is shown in Fig.\ref{fig:Fig3}(b) as a function of the control Rabi frequency $\Omega$ and the blockaded optical depth $OD_b$. In the strong-driving limit, $\Omega\gg\gamma_p$, where the entanglement-generation rate is highest, we find a simple power-law scaling $l^\star/r_b\sim OD_b^{0.45}$. This behaviour follows that of the characteristic length scale $r_h\equiv r_b\sqrt{OD_b}$ \cite{thompson2017symmetry,khazali2019polariton,supplementary} at which the dipolar excitation-exchange is most efficient [see Fig.~\ref{fig:Fig3}(a)]. In particular, $l^{\star}>r_b$ for $OD_b>1$, which suppresses interaction-induced photon losses.

In Figure~\ref{fig:Fig3}(c) we show the obtained  entanglement rate $\Gamma_{\rm e}^\star$ for the optimal pulse length and geometry as a function of $OD_b$ and  $\Omega$. For a fixed value of $OD_b$, $\Gamma_{\rm e}^\star$ increases with $\Omega$ but eventually saturates to a constant value as one enters the reflection-limited regime, where $B_R<B_T$. In this limit -- where the entanglement generation is fastest -- we find a simple power-law dependence $\Gamma_{\rm e}^\star\sim OD_b^{3/2}$ on the blockaded optical depth of the ensembles. 
We can understand this behavior from the spectral features of the propagation coefficients $\chi(\omega)$ and $\mathcal{V}(\omega)$ in Eq.(\ref{eq:MainResponse}). Analysing their corresponding expressions (cf.  \cite{supplementary}), one finds that, in the strong-driving limit and for a fixed value of the interaction $V(r)$, $\mathcal{V}$ develops two pronounced maxima at $\omega_0\sim\pm\Omega^2/V(r)$. With the maximum interaction of $V(l^\star)=\frac{\Omega^2}{\gamma_p}(r_b/l^\star)^3$, this gives $\omega_0/\gamma_p\sim(l^\star/r_b)^3\sim OD_b^{3/2}$. The separation of these maxima determines the bandwidth of $R(\omega)$ [cf. Fig.\ref{fig:Fig3} (e)], which limits the entanglement generation time $t_{\rm e}$ and, therefore,  explains the observed $\Gamma_{\rm e}^{\star}\sim OD_b^{3/2}$ power-law scaling. 
This favorable dependence on $OD_b$ is a major finding of this work and suggests that rapid entanglement generation can be achieved at sufficiently high atomic densities and Rabi frequencies of the Rydberg excitation lasers.

\begin{figure}
    \centering
    \includegraphics[width=0.95\linewidth]{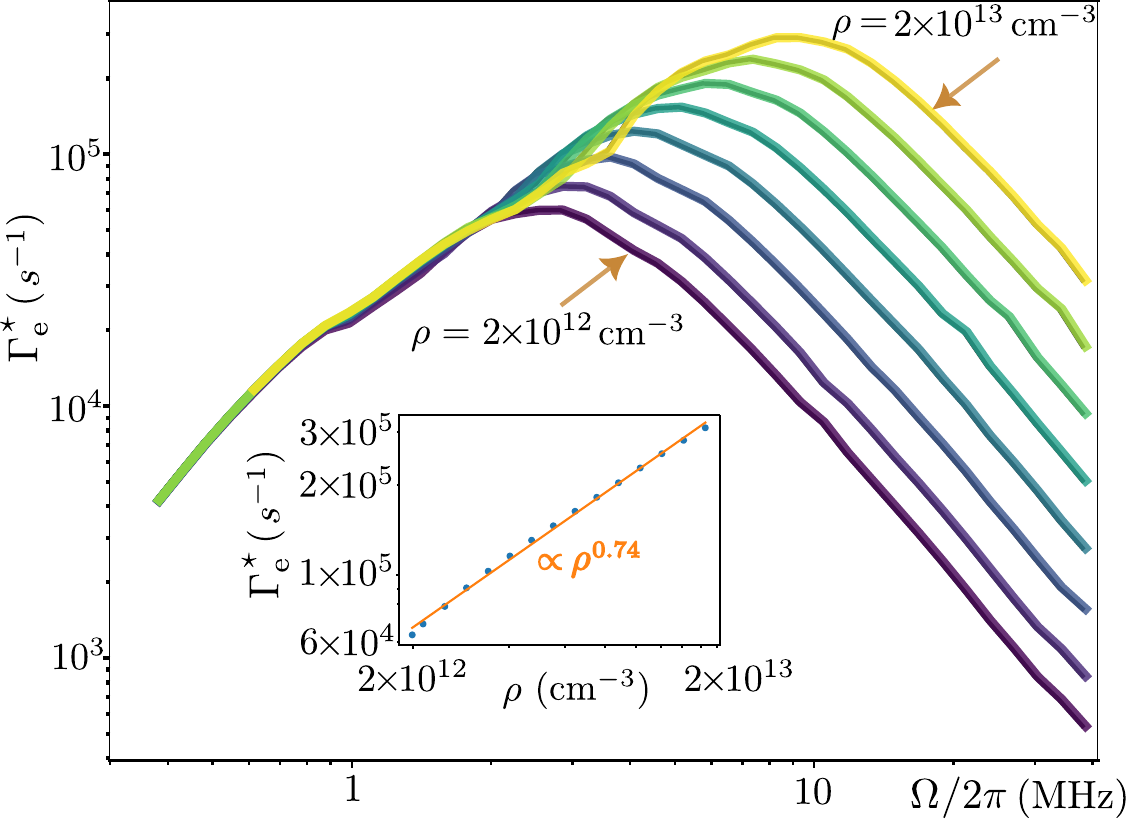}
    \caption{Optimal entanglement generation rate $\Gamma_{\rm e}^\star$ for $^{174}$Yb atoms as a function of the control Rabi frequency $\Omega$. The different curves are obtained for varying atomic peak densities between $2\times10^{12}$\,cm$^{-3}$ and $2\times10^{13}$\,cm$^{-3}$ in 7 steps on a logarithmic scale. $\Gamma_{\rm e}^\star$ features a pronounced maximum, which is shown in the inset.}
    \label{fig:Fig4}
\end{figure}

We further quantify the entanglement generation rate by considering an ensemble of Ytterbium-174 atoms \cite{takasu2003spin,saskin2019narrow,jenkins2022ytterbium,wilson2019trapped} coupled to a stationary Ytterbium-171 atom \cite{Wilson2022,muniz2025high, peper2025spectroscopy}. 
Choosing $^{174}$Yb is particularly well suited for the present protocol because its narrow intercombination $^1S_0-^3\hspace{-1.5mm}P_1$ transition, with $\gamma_p/2\pi=91$ kHz \cite{peper2025spectroscopy}, enables operation in the strong-driving regime of $\Omega\gg\gamma_p$ with lower control-field power. In addition, it exhibits favorable, balanced Clebsch-Gordan coefficients for the coupling of the two photon modes $\hat{\mathcal{E}}_\pm$. The specific choice of atomic levels and excitation schemes is detailed in the End Matter section and illustrated in Fig.\ref{fig:EndMatter}. We consider a characteristic loss rate due to spontaneous decay and black body radiation of $\gamma_r/2\pi\sim 5\:{\rm kHz}=0.055\gamma_p$ \cite{finkelstein2021continuous,finkelstein2023practical} and use $C_3 = 20$ GHz\,$\mu$m$^3$, which is a typical interaction strength for $^{174}$Yb Rydberg states with principal quantum numbers of $n\sim100$ \cite{Wilson2022, mogerle2026accurate,supplementary}. We simulate the complete spatio-temporal dynamics for a Gaussian input pulse, describing three-dimensional pulse distortion and accounting for the anisotropy of the dipole-dipole interaction potential $V({\bf r})$ (see End Matter for details). Figure \ref{fig:Fig3} shows the obtained entanglement-generation rate as a function of the control-field Rabi frequency for different values of the peak atomic density. At small values of $\Omega$, $\Gamma_{\rm e}^{\star}$ increases, as discussed above, but eventually decreases when the control-field becomes too strong. This high-intensity drop of the entanglement-generation rate stems from the decrease of $OD_b$ as the blockade radius $r_b=[C_3/(\Omega^2/\gamma_p)]^{1/3}$ is decreased when raising $\Omega$ at a fixed atomic density. As shown in the inset of Fig.~\ref{fig:Fig4}, we find that the maximum  rate scales as $\Gamma_{\rm e}^\star/\gamma_p\sim \rho^{0.74}$, while the required Rabi frequency $\Omega^\star$ to reach this value increases as $\Omega^\star/\gamma_p \sim \rho^{0.57}$ with the ensemble density.
This implies that $\Gamma^{\star}_{\rm e}/\gamma_p\sim {OD_b^{\star}}^{1.19}$ increases with the blockaded optical depth $OD_b^{\star}$ at $\Omega^{\star}$.
Therefore, the high value $\Gamma_{\rm e}^\star=3\times10^5$\,s$^{-1}$ obtained for $\rho=2\times 10^{13}$\,cm$^{-3}$ can be further improved by increasing $\rho$, $C_3$, $\gamma_p$, or $l_{\rm abs}^{-1}$. Moreover, the ensemble-based approach is well suited for parallel operation of multiple ensemble-pairs using optical tweezer arrays [cf. Fig.\ref{fig:Fig1}(a)], enabling rapid entanglement generation at the speed of current neutral-atom quantum gates. 

In conclusion, we described a method for high-rate remote entanglement generation between neutral-atom quantum processors. The scheme uses the key resource of neutral-atom arrays, namely the strong dipole-dipole interaction between atomic Rydberg states and achieves strong qubit-photon coupling via qubit interactions with a mesoscopic atomic ensemble. The use of narrow-linewidth transitions, such as in $^{174}$Yb atoms, suggests that entanglement generation rates that approach the speed of current two-qubit gates are experimentally feasible. While our approach does not require optical resonators, combining atomic ensembles with low-finesse cavities \cite{Vaneecloo2022,Stolz2022,desantis2026} would enhance the performance of the presented scheme. Additionally, the approach is naturally suited for time-dependent control schemes \cite{fleischhauer2005electromagnetically} to further speed up entanglement generation. 
Given the high entanglement-generation rate of the proposed interconnect, it appears promising to explore how it can be best utilized in the design of error-corrected quantum networks \cite{munro2010quantum,sinclair2025fault}, possibly operating at telecom frequencies \cite{li2025parallelized}. The described qubit-controlled reflection offers a basic mechanism to generate single photons, as required to operate the interconnect. This might be exploited to design future protocols that use coherent light, instead of single-photon pulses, reducing the technical requirements to realize high entanglement generation rates. More generally, the qubit-controlled optical chirality of the nodes in our setup presents an interesting outlook as novel elements for waveguide-QED settings to explore exotic collective light-matter phenomena \cite{lohdal2017,sheremet2023waveguide}.  

We thank Fan Yang, Tao (Alex) Zheng, Greg
Ferrero, Jan Kumlin, Majid Zahedian, and Zhanchuan Zhang for valuable discussions.
This research was funded in whole or in part
by the Swiss State Secretariat for
Education, Research and Innovation (SERI) – Grant no.: Uem029-2-225224, the European Union’s Horizon Europe research
and innovation program under  
the European Research Council through the ERC Synergy
Grant SuperWave (Grant No. 101071882), the cluster of excellence quantA (Grant No. 10.55776/COE1) and the European Union (NextGenerationEU). 

\bibliographystyle{unsrt}
\bibliography{Interconnect}

\section{End Matter}
\begin{figure}[t!]
    \centering
\includegraphics[width=\linewidth]{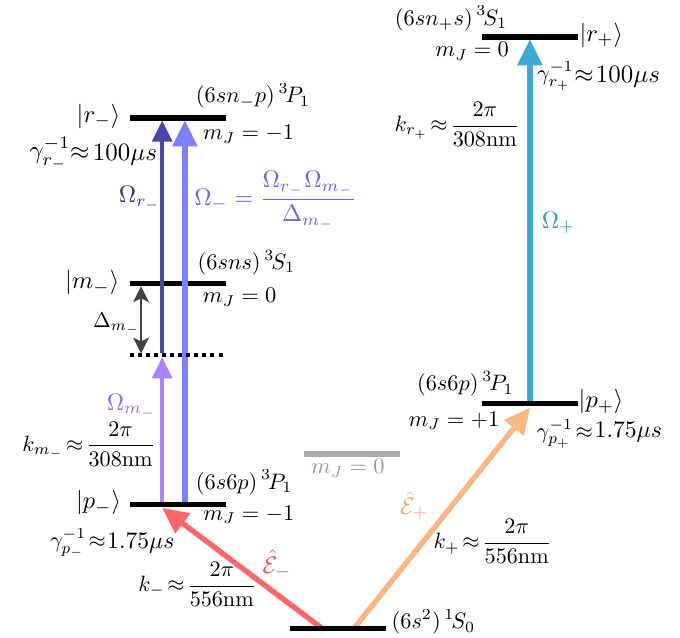}
    \caption{Energy level diagrams of the $^{174}$Yb ensemble atoms, indicating all relevant states, lifetimes, and transition wavenumbers, discussed in the text. }
    \label{fig:EndMatter}
\end{figure}
Figure~\ref{fig:EndMatter} shows the detailed level structure for the $^{174}$Yb ensemble atoms discussed above. The atoms are polarized with a strong magnetic field oriented along the $y$-axis [see Fig.~\ref{fig:Fig3}(a)], perpendicular to the distance vector between the ensemble and the qubit atom ($x$-axis, cf. Fig.~\ref{fig:EndMatter}).
The circularly polarized photon field, which propagates along the $z$-axis, therefore, drives the transition between the singlet ground state $\ket{g}=\ket{^1S_0, F=0, m_F=0}$ and the
triplet states $\ket{p_+}=\ket{^3P_1,F=1,m_F=1}$ and $\ket{p_-}=\ket{^3P_1,F=1,m_F=-1}$, with $\gamma_p = 2\pi \times 91$ kHz. 
The two intercombination-line transitions have identical Clebsch-Gordan coefficients and, thus, yield equal values of $OD_b$ for the forward and backward propagating photon mode. The control-laser fields couple the $\lvert p_\pm\rangle$-states to the opposite-parity Rydberg states $\lvert r_+\rangle=\lvert(6sn_+ s)^3S_1,m_J=0\rangle$ and $\lvert r_-\rangle=\lvert(6sn_- p)^3P_1,m_J=-1\rangle$. For the stationary qubit we use a $^{171}$Yb atom, with the singlet  states $\ket{0}=\ket{(6s^2)^1S_0,m_F=-1/2}$ and $\ket{1}=\ket{(6s^2)^1S_0,m_F=+1/2}$ as the ground state qubit, and $\ket{\downarrow}=\ket{(6sn_{\downarrow}p)^1P_1,m_F=3/2}$ and $\ket{\uparrow}=\ket{(6sn_{\uparrow}p)^1S_0,m_F=1/2}$ to encode the qubit in the Rydberg manifold \cite{supplementary}.

In order to reach the Rydberg-state $\ket{r_-}$ in the $(6sn_-s)^3P_1$ manifold, we use a resonant two-photon transition with an intermediate Rydberg state $\ket{m_-}$ that is excited by an optical field with Rabi frequency $\Omega_{m_-}$ and detuning $\Delta_{m_-}$. The secondary Rydberg transition is driven by a microwave field with Rabi frequency $\Omega_{r_-}$, close to two-photon resonance. For sufficiently large $\Delta_{m_-}$, the coupling of the $\ket{p_-}-\ket{r_-}$ transition can then be described by the two-photon Rabi frequency $\Omega_-=\frac{\Omega_{r_-}\Omega_{m_-}}{\Delta_{m_-}}$, which we use in the main text. We use a  decay rate of $\gamma_{r}=2\pi\times 5$ kHz for the two Rydberg states, corresponding to typical experimental lifetimes \cite{peper2025spectroscopy}.

For this configuration and sufficient Zeeman splitting between the magnetic sub-levels, the interaction strength in Eq.~(\ref{eq:Interactions}) can be written as \cite{supplementary}
\begin{align}
    V(\mathbf{r} ) = \frac{C_3}{r^3}\sin^2(\theta) ,
    \label{eq:Angular}
\end{align}
where $\theta$ is the angle between the magnetic-field axis ($y$-axis) and the interatomic distance vector. 

The interaction is maximal for $\theta=90^{\circ}$, i.e. for ensemble atoms in the $z-x$ plane. Consequently, smaller beam waists yield a larger entanglement rate. Since we do not account for transverse diffraction in our calculations, we choose the RMS radius of the Gaussian input mode as $\sigma = \sqrt{\frac{L^{\star} \lambda}{\pi}}$, in order to ensure negligible diffraction effects. In cases where the resulting $\sigma$ is comparable to the optimized qubit-ensemble distance $l^{\star}$, we reduce the ensemble length $L$ to satisfy $\sqrt{\frac{L  \lambda}{\pi}}=\sigma \le l^{\star}/9$ such that the qubit atom remains well separated from the photon mode. 
\end{document}